\documentclass[12pt]{article}

\ifx\pdfoutput\undefined
\usepackage[dvips,bookmarks]{hyperref}
\else
\usepackage{hyperref}
\fi
\hypersetup{colorlinks=false,bookmarksopen,bookmarksnumbered,citecolor=blue,
   pdfstartview=FitH}

\usepackage{latexsym}
\usepackage{amssymb,amsfonts,amsmath}
\usepackage{graphicx} 
\usepackage{indentfirst}
\usepackage{bbm}

\parskip=\medskipamount

\newcommand {\cD}{{\cal D}}

\newcommand {\cF}{{\cal F}}

\newcommand {\cL}{{\cal L}}
\newcommand {\cM}{{\cal M}}
\newcommand {\cN}{{\cal N}}
\newcommand {\cO}{{\cal O}}

\newcommand {\cV}{{\cal V}}

\newcommand {\cX}{{\cal X}}

\newcommand {\cZ}{{\cal Z}}

\def\a{\alpha}
\def\b{\beta}
\def\c{\chi}
\def\d{\delta}

\def\g{\gamma}

\def\l{\lambda}

\def\q{\theta}

\def\s{\sigma}

\def\z{\zeta}
\def\D{\Delta}
\def\F{\Phi}
\def\J{\Psi}
\def\L{\Lambda}

\def \bi{\bibitem}

\def\ri{{\rm i}}
\def\re{{\rm e}}

\newcommand{\ad}{{\dot{\alpha}}}                           
\newcommand{\bd}{{\dot{\beta}}}                            
\newcommand{\ve}{\varepsilon}                            

\newcommand{\pa}{\partial}                           
\newcommand{\hf}{\frac12}

\newcommand{\be}{\begin{equation}}
\newcommand{\ee}{\end{equation}}
\newcommand{\bea}{\begin{eqnarray}}
\newcommand{\eea}{\end{eqnarray}}
\newcommand{\non}{\nonumber}
\newcommand{\ba}{\begin{array}}
\newcommand{\ea}{\end{array}}

\newcommand{\bm}[1]{\mbox{\boldmath$#1$}}

\def\double #1{#1{\hbox{\kern-2pt $#1$}}}

\newcommand{\ha}{{\hat{a}}}
\newcommand{\hb}{{\hat{b}}}
\newcommand{\hc}{{\hat{c}}}
\newcommand{\hd}{{\hat{d}}}
\newcommand{\he}{{\hat{e}}}
\newcommand{\hM}{{\hat{M}}}

\newcommand{\hA}{{\hat{A}}}
\newcommand{\hB}{{\hat{B}}}
\newcommand{\hC}{{\hat{C}}}
\newcommand{\hD}{{\hat{D}}}
\newcommand{\hE}{{\hat{E}}}
\newcommand{\hF}{{\hat{F}}}

\newcommand{\gd}{{\dot\g}}
\newcommand{\dd}{{\dot\d}}

\newcommand{\ts}{{\tilde{\s}}}

\newcommand{\bsubeq}{\begin{subequations}}
\newcommand{\esubeq}{\end{subequations}}

\newcommand{\ul}{\underline}

\newcommand{\dalpha}{{\dot{\alpha}}}
\newcommand{\dbeta}{{\dot{\beta}}}
\newcommand{\dgamma}{{\dot{\gamma}}}
\newcommand{\ddelta}{{\dot{\delta}}}
\newcommand{\dmu}{{\dot{\mu}}}

\newcommand{\eol}{\notag \\}
\newcommand{\rd}{\mathrm d}
\newcommand{\HC}{{\mathrm{c.c.}}}
\newcommand{\veps}{\varepsilon}

\newcommand{\halpha}{{\hat \alpha}}
\newcommand{\hbeta}{{\hat \beta}}
\newcommand{\hgamma}{{\hat \gamma}}
\newcommand{\hdelta}{{\hat \delta}}

\newcommand{\bnabla}{{\bm\nabla}}

\begin{document}

\begin{titlepage}
\begin{flushright}
May 2012\\
\end{flushright}
\vspace{5mm}

\begin{center}
{\Large \bf The linear multiplet and ectoplasm}
\\ 
\end{center}

\begin{center}

{\bf Daniel Butter, Sergei M. Kuzenko and Joseph Novak}

\footnotesize{
{\it School of Physics M013, The University of Western Australia\\
35 Stirling Highway, Crawley W.A. 6009, Australia}}  ~\\
\texttt{daniel.butter,\,sergei.kuzenko,\,joseph.novak@uwa.edu.au}\\
\vspace{2mm}

\end{center}
\vspace{5mm}

\begin{abstract}
\baselineskip=14pt
In the framework of the superconformal tensor calculus for 4D $\cN=2$ supergravity,
locally supersymmetric actions are often constructed using the linear multiplet.  
We provide a superform formulation for the linear multiplet and derive  the corresponding action functional using the ectoplasm method (also known as  the superform approach to the 
construction of supersymmetric invariants). 
We propose a  new locally supersymmetric action which makes use of a deformed linear multiplet. The novel feature of this multiplet is that it corresponds to the case of a gauged  central charge using a one-form potential not annihilated by the central charge (unlike the standard $\cN=2$ vector multiplet). Such a gauge one-form can be chosen to describe a variant nonlinear vector-tensor multiplet. 
As a byproduct of our construction, we also find a variant realization of the
tensor multiplet in supergravity where one of the
auxiliaries is replaced by the field strength of a gauge three-form.
\end{abstract}
\vspace{0.5cm}

\vfill
\end{titlepage}

\newpage
\renewcommand{\thefootnote}{\arabic{footnote}}
\setcounter{footnote}{0}



\numberwithin{equation}{section}



\newpage
\section{Introduction}

In $\cN=2$ Poincar\'e supersymmetry in four space-time dimensions (4D), 
the linear multiplet was introduced by Sohnius  \cite{Sohnius} as a superfield Lagrangian 
describing the dynamics of matter hypermultiplets coupled to Yang-Mills superfields 
\cite{GSW}. Following \cite{Sohnius,BS}, the linear multiplet 
is a real isotriplet superfield, $L^{ij}=L^{ji}$
and  $\overline{L^{ij}}=L_{ij} := \ve_{ik} \ve_{jl} L^{kl}$, 
 subject to the constraints
\bea 
D_\a^{(i} L^{jk)} = \bar D_\ad^{(i} L^{jk)} = 0 ~ .
\eea
Here $D^i_\a$ and $\bar D^i_\ad$ are the $\cN=2$ spinor covariant derivatives with a 
real central charge $\D$.
The action proposed in  \cite{Sohnius} has the form 
\bea 
S = - \frac{1}{12} \int \rd^4 x\,  \Big( D^{\a (i} D_\a^{j)}
+  \bar D_\ad^{ (i} \bar D^{  j) \ad}
\Big) L_{ij } \Big|_{\q=0}   ~.
\label{1.2}
\eea
It is invariant under the $\cN=2$ super-Poincar\'e transformations, 
including the central charge one.
The name `linear multiplet' was coined by Breitenlohner and Sohnius \cite{BS}
because  the decomposition of $L^{ij} $
into $\cN=1$ superfields  contains a real linear multiplet \cite{FWZ} (which is the field strength 
of the $\cN=1$ tensor multiplet \cite{Siegel})
in the case that $L^{ij}$ is neutral under the central charge,
$\D L^{ij}=0$. Unlike the hypermultiplet, demanding $\D L^{ij}=0$ does not lead to an on-shell 
multiplet. The resulting off-shell multiplet without central charge  
\cite{SSW2} is naturally  interpreted as the field strength of 
the massless $\cN=2$ tensor multiplet \cite{Wess}.

The action \eqref{1.2} may be thought of as an $\cN=2$ analogue of the chiral action in $\cN=1$
supersymmetry. As is well known, any $\cN=1$ action can be rewritten as  a chiral one.
The situation in $\cN=2$ supersymmetry is similar. 
As stated by Breitenlohner and Sohnius \cite{BS}, all known Lagrangians ({\it at that time})
for rigid $\cN=2$ supersymmetry can be generated from linear multiplets.  
Since the linear multiplet was lifted to $\cN=2$ supergravity \cite{BS}, and then 
reformulated  \cite{deWvHVP3} within the $\cN=2$ superconfomal tensor calculus
\cite{deWvHVP,BdeRdeW,deWvHVP2}, it has become a universal tool to construct 
the component actions for supergravity-matter systems, 
especially within the locally superconformal setting of \cite{deWvHVP,BdeRdeW,deWvHVP2}. 

In regard to the superspace practitioners, for a long time they had not expressed much interest
in the linear multiplet, since there had appeared  more powerful methods to construct 
off-shell supersymmetric actions using the harmonic \cite{GIKOS,GIOS} and 
the projective \cite{KLR,LR} superspace approaches which are based 
on the use of superspace  ${\mathbb R}^{4|8}\times {\mathbb C}P^1 $ 
pioneered by Rosly \cite{Rosly}. 
The situation changed in the mid-1990s when 
the so-called vector-tensor multiplet \cite{SSW} 
was re-discovered  by string theorists \cite{deWKLL} to be important in the context of string 
compactifications. This multiplet is analogous 
to the Fayet-Sohnius multiplet \cite{Fayet,Sohnius} in the sense that it possesses
an intrinsic central charge (i.e. the multiplet is on-shell if the central charge vanishes),
and therefore its dynamics (including its couplings to  vector multiplets and supergravity) 
should be described by a linear multiplet Lagrangian. 
The vector-tensor multiplet and its nonlinear version \cite{Claus1,Claus2}
have become the subject of various studies in flat superspace
 \cite{HOW,GHH,DKT,BHO,DK,IS,DIKST}. In particular, a general harmonic superspace formalism 
for 4D $\cN=2$ rigid supersymmetric theories with gauged central charge was developed in \cite{DIKST}. Furthermore,  a remarkable construction was given by Theis \cite{Theis1,Theis2}. 
He proposed a new nonlinear 
vector-tensor multiplet with the defining properly that the central charge is gauged using the vector 
field belonging to the multiplet (unlike the approach of \cite{DIKST} which used an off-shell vector multiplet 
to gauge the central charge). 

The action \eqref{1.2} can be represented as a superspace integral \cite{DKT}, 
but this requires, in the case $\D L^{ij}\neq 0$, the use of harmonic superspace 
\cite{GIKOS}.
Introducing SU(2) harmonics $u^{+i}$ and $u^-_i$ according to \cite{GIKOS}, 
one can associate with $L^{ij}$ the following {\it analytic} superfield 
$L^{++} := u^+_iu^+_j L^{ij}$ which is annihilated by $D^+_\a := u^+_i D^i_\a$ and 
$ \bar D^+_\ad := u^+_i \bar D^i_\ad$.
Then, the action \eqref{1.2} is equivalent to
\bea
S= \int    \rd u \,\rd \z^{(-4)}\, \Big( \q^{+\a} \q^+_\a + \bar \q^+_\ad \bar \q^{+\ad}\Big) L^{++}~,
\label{1.3}
\eea
where $\q^+_\a := u^+_i \q^i_\a$ and $\bar \q^+_\ad := u^+_i \bar \q^i_\ad$.
The integration in \eqref{1.3} is carried over the analytic subspace of the harmonic superspace. 
In particular, 
$\rd u$ denotes the left-right invariant measure of SU(2),  and  
\bea
\rd \z^{(-4)} = \frac{1}{16} \rd^4 x D^{- \a} D^-_\a  \bar D^{-}_\ad \bar D^{-\ad} ~, 
\qquad  D^-_\a := u^-_i D^i_\a~, \quad \bar D^-_\ad := u^-_i \bar D^i_\ad~.
\eea
The supergravity extension of \eqref{1.3} was given in \cite{KT}
\bea
S_{\rm SUGRA}= \int \rd u\,\rd \z^{(-4)} \, \cV_5^{++} \cL^{++}~.
\label{1.5}
\eea
Here $\cL^{++}$ denotes the curved-superspace version of the linear multiplet, 
while $\cV^{++}_5$ is the harmonic prepotential of one of the two supergravity compensators --
a vector multiplet which gauges the central charge. 
The action \eqref{1.5} is a locally supersymmetric extension of the action introduced in \cite{DIKST}.
The combination $V_5^{++} :=  (\q^{+} )^2 + (\bar \q^+)^2$ in 
\eqref{1.3} can be interpreted as the analytic prepotential of a frozen vector multiplet with 
constant field strength \cite{DIKST}. 
The functional \eqref{1.5} is extremely compact and geometric, as compared with 
its component counterpart  \cite{deWvHVP3} (see eq. \eqref{eq_LinearAction} below).
Remarkably,  $S_{\rm SUGRA}$ is a $BF$-type model invariant under 
 gauge transformations of the form \cite{KT}: 
\bea
\d \cV^{++}_5 = - \cD^{++}\l~, \qquad \d \cL^{++} = \l \,\D \cL^{++}~,
\eea
with $\l$ an analytic gauge parameter, and $\cD^{++}$ a harmonic gauge-covariant derivative 
defined in \cite{KT}. 
Unfortunately, the above action is  not yet useful for practical applications.  
The point is that the harmonic superspace formulation of $\cN= 2$ supergravity 
was developed in terms of certain prepotentials \cite{Galperin:1987em,Galperin:1987ek} 
(see also \cite{GIOS} for a review). It is not known how to derive the prepotential 
description of \cite{Galperin:1987em,Galperin:1987ek}  
from the three existing superspace formulations  for 4D $\cN=2$ conformal supergravity 
\cite{Howe,KLRT-M1,Butter4D}.\footnote{As shown in   \cite{KLRT-M2}, 
the formulation developed in \cite{KLRT-M1} can be obtained from \cite{Howe}
by a partial gauge fixing of the super-Weyl invariance. The latter formulation 
is a gauged-fixed version of the conformal supergravity formulation developed in 
\cite{Butter4D}.  One can think of the formulation \cite{Butter4D} as a master one.
Depending on a concrete application, it is convenient to use 
either \cite{KLRT-M1} or \cite{Butter4D}.}
These formulations are realized in terms of covariant derivatives defined on a curved 
$\cN=2$ superspace. The difference between the three formulations 
 lies in the structure groups chosen. What is important is that 
 all known multiplets with gauged central charge in the presence of supergravity 
are realized in curved superspace in terms of the supergravity covariant derivatives 
\cite{KN,BN}, and not in terms of the harmonic prepotentials. 
Therefore, we need a reformulation of the linear multiplet action \eqref{1.5} that is
given solely in terms of the supergravity covariant derivatives. Such a reformulation
is given in the present paper.

Our work contains two main results. Firstly, we develop a superform formulation for 
the linear multiplet in $\cN=2$ conformal supergravity. This formulation is shown 
to immediately lead to a locally supersymmetric action if we make use of the so-called 
ectoplasm formalism  \cite{Ectoplasm,GGKS}
(also known as  the superform approach to the 
construction of supersymmetric invariants).\footnote{The mathematical construction 
underlying the  ectoplasm formalism  \cite{Ectoplasm,GGKS} is a special case of 
the theory of integration over surfaces in supermanifolds, see \cite{Vor} and references  
therein. In the physics literature, the idea to use closed super
four-forms for the construction of locally supersymmetric actions in 4D
was, to the best of our knowledge, first given by Hasler \cite{Hasler}
building on the analysis in \cite{GG}. }
The action derived coincides with that 
introduced in \cite{deWvHVP3}. Secondly, we propose a  new locally supersymmetric action which makes use of a deformed linear multiplet. The novel feature of this multiplet is that it corresponds to the case of the central charge being gauged using a one-form potential which is not annihilated by the central charge 
(unlike the standard $\cN=2$ vector multiplet). 

This paper is organized as follows. Section 2 describes a warm-up construction.  
We start from a superform realization for the linear multiplet without a central charge
in 5D $\cN=1$ Minkowski superspace, and use it to read off a superform formulation for 
the linear multiplet in flat 4D $\cN=2$ central charge superspace. 
In section 3 we provide a superform formulation for the linear multiplet 
in $\cN=2$ conformal supergravity and derive  the corresponding action functional 
using the ectoplasm method. In section 4 we first review, following \cite{Novak},
the curved-superspace formulation for a generalized $\cN=2$ vector multiplet 
which gauges the central charge and is not inert under the central charge transformations 
(unlike the standard $\cN=2$ vector multiplet). We then develop a superform formulation 
for a deformed linear multiplet and construct the associated locally supersymmetric action. 
The main body of the paper is accompanied by two appendices.
The first appendix is technical and devoted to a brief summary 
of the superspace formulation for $\cN=2$ conformal supergravity 
developed in  \cite{Butter4D} and slightly reformulated in \cite{BN}.
The second appendix briefly describes the ectoplasm formulation
of the $BF$ coupling in $\cN=1$ conformal supergravity.

\section{The linear multiplet in flat superspace}
In this section, we briefly discuss the linear multiplet $L^{ij}$ in flat
superspace and describe its superform structure. It is well known that the
linear multiplet in 4D with a central charge is related to a linear multiplet
in 5D without a central charge.\footnote{In 5D, a linear multiplet without
central charge also has been called a tensor multiplet or an $\cO(2)$ multiplet
in the literature, in analogy to the terminology used in 4D.}
We will first describe the situation in 5D and then demonstrate its
equivalence to the 4D case with a central charge.

\subsection{The linear multiplet in flat 5D superspace}
We use the 5D superspace and gamma matrix conventions of \cite{KL}, 
to which we refer the reader.
The algebra of 5D flat covariant derivatives\footnote{The 5D flat covariant derivatives are 
$D_\hA = ( \pa_\ha , D_{\underline{\halpha}}) $, where 
$D_{\underline{\halpha} }:= D_\halpha^i$. The dual basis of one-forms is 
$E^\hA = ( E^\ha , E^{\underline{\halpha}}) $, where 
$E^{\underline{\halpha} }:= E^\halpha_i$.} 
is
\begin{align}
\{D_\halpha^i, D_\hbeta^j\} = -2\ri \,\ve^{ij}
(\Gamma^\hc)_{\halpha \hbeta}  \pa_{\hc}~, \qquad
[D_\halpha^i, \pa_\hb] = 0~, \qquad [\pa_\ha, \pa_\hb] = 0~.
\end{align}
The linear multiplet in 5D is encoded in a real
linear superfield $L^{ij} = (L_{ij})^*$ which is
symmetric in its indices, $L^{ij} = L^{ji}$, and obeys
the constraints
\begin{align}\label{eq_Glin}
D_\halpha^{(i} L^{jk)} = 0~.
\end{align}
These constraints imply the existence of a conserved vector among the components
of $L^{ij}$,
\begin{align}
V^\ha = \frac{\ri}{24} (\Gamma^\ha)^{\halpha \hbeta} D_{\halpha j} D_{\hbeta k} L^{jk}\vert~, \qquad
\pa_{\ha} V^\ha = 0~.
\end{align}
The conserved vector $V^\ha$ is naturally dual to a closed four-form.

It is useful to introduce a superspace generalization of this four-form so that
the linearity constraint \eqref{eq_Glin} appears naturally as a Bianchi identity.
Let $\widehat\Sigma$ be a closed four-form\footnote{We place a hat on $\Sigma$ to
distinguish it from the four-form $\Sigma$ in 4D that we will introduce in the next
section.} with a tangent frame expansion
\begin{align}
\widehat\Sigma = \frac{1}{4!} E^\hD \wedge E^\hC \wedge E^\hB \wedge E^\hA \, \widehat\Sigma_{\hA \hB \hC \hD}~.
\end{align}
The requirement that $\widehat\Sigma$ is closed, $\rd \widehat\Sigma = 0$, 
amounts to the equations
\begin{align}\label{eq_dSigma}
0 = D_{[\hA} \widehat\Sigma_{\hB\hC\hD\hE\}}
	- 2 T_{[\hA \hB|}{}^{\hF} \widehat\Sigma_{\hF| \hC \hD \hE\}}~,
\end{align}
where the indices $\hA \cdots \hE$ are {\it graded} anti-symmetrized.
Imposing the constraints
\begin{align}
\widehat\Sigma_{\ul {\halpha\hbeta\hgamma\hdelta}} = 0~, \qquad
\widehat\Sigma_{\ha \ul{\hbeta \hgamma \hdelta}} = 0~, \qquad
\widehat\Sigma_{\ha \hb}{}_\halpha^i{}_\hbeta^j = 4\ri (\Sigma_{\ha \hb})_{\halpha \hbeta} L^{ij}~,
\end{align}
for some real symmetric tensor $L^{ij}$, we find that the Bianchi identities
require \eqref{eq_Glin} and fix the remaining components of the
four-form:
\begin{align}
\widehat\Sigma_{\ha\hb\hc}{}_\halpha^i = -\frac{1}{3} \veps_{\ha\hb\hc\hd\he} (\Sigma^{\hd\he})_\halpha{}^\hbeta D_{\hbeta j} L^{ji}~, \qquad
\widehat\Sigma_{\ha\hb\hc\hd} = \frac{\ri}{24} \veps_{\ha\hb\hc\hd\he} (\Gamma^\he)^{\halpha \hbeta} D_{\halpha j} D_{\hbeta k} L^{jk}~.
\end{align}
The highest component $\widehat\Sigma_{\ha\hb\hc\hd}$ is closed by construction, as a consequence of the linearity constraint \eqref{eq_Glin}.
This closed four-form has recently appeared in the literature \cite{HPSS-C}.  

It is possible to require that $\widehat\Sigma$ be an exact form, $\widehat\Sigma = \rd \widehat C$, 
for some three-form $\widehat C$. In the tangent frame,
\begin{align}
\widehat\Sigma_{\hA \hB \hC \hD} = 4 D_{[\hA} \widehat C_{\hB \hC \hD\}}
	- 6 T_{[\hA \hB|}{}^{\hE} \widehat C_{\hE |\hC \hD\}}~.
\end{align}
Then the highest component $\widehat \Sigma_{\ha\hb\hc\hd}$ is similarly exact.

\subsection{The linear multiplet in flat 4D central charge superspace}
The 5D derivatives can be decomposed into 4D $\cN=2$ derivatives
$D_A = (D_\alpha^i, \bar D^\dalpha_i, \pa_a)$ and a central
charge $\pa_5$,
\begin{align}
D_\halpha^i = (D_\alpha^i, \bar D^{\dalpha i})~, \qquad
\pa_{\ha} = (\pa_a, \pa_5)~,
\end{align}
so that the supersymmetry algebra becomes
\begin{gather}
\{D_\alpha^i, D_\beta^j\} = 2 \ve^{ij} \ve_{\alpha \beta} \pa_5~, \quad
\{\bar D_{\dalpha i}, \bar D_{\dbeta j}\} = 2 \ve_{ij} \ve_{\dalpha \dbeta} \pa_5~, \quad
[D_\alpha^i, \bar D_{\dbeta j}] = -2\ri \,\delta^i{}_j (\sigma^c)_{\alpha \dbeta} \pa_c~, \non\\
[D_\alpha^i, \pa_b] = [\pa_a, \pa_b] = [D_\alpha^i, \pa_5] = [\pa_a, \pa_5] = 0~.
\end{gather}
Any multiplet in flat 5D $\cN=1$ superspace can naturally be written in
4D $\cN=2$ superspace with a real central charge $\Delta = \pa_5$.
The linear multiplet $L^{ij}$, for example, now obeys
$D_\alpha^{(i} L^{jk)} = \bar D_\dalpha^{(i} L^{jk)} = 0$.
Its associated four-form multiplet $\Sigma_{\hA\hB\hC\hD}$ naturally decomposes into a
four-form $\Sigma_{ABCD}$ and a
three-form $H_{ABC} = \Sigma_{5 ABC}$, which are related by the 4D version
of eq. \eqref{eq_dSigma},
\begin{align}
0 &= D_{[A} \Sigma_{BCDE\}}
	- 2 T_{[A B|}{}^{F} \Sigma_{F| C D E\}}
	- 2 T_{[A B|}{}^{5} H_{|C D E\}} ~,\\
0 &= \pa_5 \Sigma_{ABCD}
	- 4 D_{[A} H_{BCD\}}
	+ 2 T_{[A B|}{}^{E} H_{E| C D\}}~.
\end{align}
The 5D torsion $T^5 := E^B \wedge E^A T_{AB}{}^5$ can be interpreted as the field strength
of a frozen vector multiplet associated
with the central charge. In form notation, these equations become
\begin{align}\label{eq_HSigmaBianchiFlat}
D \Sigma = H \wedge T^5~, \qquad
D H = \pa_5 \Sigma~, \qquad D := E^A D_A~,
\end{align}
where $D$ is the central charge covariant exterior derivative, obeying $D^2 = T^5 \pa_5$.
The three-form $H$ has the components
\begin{subequations}\label{eq_Hcompflat}
\begin{gather}
H_{\ul{\alpha \beta \gamma}} = H_{\ul{\alpha \beta \dgamma}} = 0~, \qquad
H_a{}_\beta^i{}_\gamma^j = H_a{}_\dbeta^i{}_\dgamma^j = 0~, \qquad
H_a{}_\beta^i{}_\dgamma^j = 2 (\sigma_a)_{\beta \dgamma} L^{ij}~, \\
H_{ab}{}_\alpha^i = -\frac{2\ri}{3} (\sigma_{ab})_\alpha{}^\beta D_{\beta j} L^{ji}~, \qquad
H_{ab}{}^\dalpha_i = -\frac{2\ri}{3} (\tilde\sigma_{ab})^\dalpha{}_\dbeta \bar D^{\dbeta j} L_{ji}~, \\
H_{abc} = \frac{\ri}{24} \ve_{abcd} (\tilde\sigma^d)^{\dbeta \alpha} [D_{\alpha i}, \bar D_{\dbeta j}] L^{ij}~,
\end{gather}
\end{subequations}
and the four-form $\Sigma$ is\footnote{In 4D, we use a hatted spinor index
to denote a four component spinor, \it{e.g.} $\psi_{\hat \alpha} = (\psi_\alpha, \bar \psi^{\dalpha})$.}
\begin{subequations}\label{eq_Scompflat}
\begin{gather}
\Sigma_{\ul {\halpha\hbeta\hgamma\hdelta}} = 0~, \qquad
\Sigma_{a \ul{\hbeta \hgamma \hdelta}} = 0~, \qquad
\Sigma_{a b}{}_\alpha^i{}_\dbeta^j = 0~, \qquad
\Sigma_{a b}{}_\alpha^i{}_\beta^j = 4\ri (\sigma_{ab})_{\alpha \beta} L^{ij}~, \\
\Sigma_{abc}{}_\alpha^i = -\frac{\ri}{3} \veps_{abcd} (\sigma^d)_{\alpha\dbeta} \bar D^{\dbeta}_j L^{ji}~, \qquad
\Sigma_{abcd} = \frac{1}{24} \veps_{abcd} (D_{ij} L^{ij} + \bar D_{ij} L^{ij})~,
\end{gather}
\end{subequations}
where $D^{ij} := D^{\alpha (i} D_\alpha^{j)}$ and $\bar D^{ij} := \bar D_{\dalpha}^{(i} \bar D^{\dalpha j)}$.
One can check that the closure condition $(DH)_{abcd} = \pa_5 \Sigma_{abcd}$ amounts to
\begin{align}\label{eq_4DClosedVector}
-\ri \pa^{\dbeta \alpha} [D_\alpha^i, \bar D_\dbeta^j] L_{ij} = \pa_5 (D^{ij} L_{ij} + \bar D^{ij} L_{ij})~.
\end{align}

What meaning can we give to these forms?
The highest component of the four-form, $\Sigma_{abcd}$, has an immediate
physical interpretation: it is the Sohnius Lagrangian \eqref{1.2}, which
associates to any linear multiplet $L^{ij}$ a supersymmetric action
principle.

The meaning of $H$, on the other hand, is clearest if we restrict
to the case where $L^{ij}$ is independent of the central charge, $\pa_5 L^{ij} = 0$.
Then the linear multiplet becomes a tensor multiplet.
In this case $H$ is a closed three-form $D H = \rd H = 0$, and its components
\eqref{eq_Hcompflat} coincide with the usual encoding of a tensor multiplet
into a closed three-form geometry. In particular, $H_{abc}$ is a closed three-form
and dual to a conserved vector since the right-hand side of \eqref{eq_4DClosedVector}
vanishes. In this case, $H$ is usually interpreted as the field strength
of a two-form $B$.

Just as in 5D, we may restrict to the case where these superforms are exact.
The three-form potential $\widehat C$ in 5D decomposes in 4D into a three-form
potential $C$ and a two-form $B$,
\begin{align}
C_{ABC} = \widehat C_{ABC}~, \qquad B_{AB} = - \widehat C_{5 AB}~,
\end{align}
so that $H$ and $\Sigma$ are given respectively by
\begin{subequations}
\begin{align}
H_{ABC} &= 3 D_{[A} B_{BC\}} - 3 T_{[AB|}{}^D B_{D|C\}} + \Delta C_{ABC}~,\\
\Sigma_{ABCD} &= 4 D_{[A} C_{BCD\}} - 6 T_{[AB|}{}^E C_{E|CD\}} + 6 T_{[AB}{}^5 B_{CD\}}~,
\end{align}
\end{subequations}
or, equivalently,
\begin{align}
H = D B + \Delta C~, \qquad \Sigma = D C + B \wedge T^5~.
\end{align}
These equations automatically satisfy the Bianchi identities \eqref{eq_HSigmaBianchiFlat}.
An interesting consequence of the exactness condition is that one of the auxiliary
components of the linear multiplet becomes the dual of a four-form field strength,
\begin{align}\label{eq_LAux}
D_{ij} L^{ij} + \bar D_{ij} L^{ij} = -4 \veps^{abcd} \pa_{a} C_{bcd}~.
\end{align}
This equation holds even in the absence of a central charge, where it describes
a variant representation of the 4D $\cN=2$  tensor multiplet 
obtained from the latter by replacing one of its auxiliary scalars by 
the field strength of a gauge three-form
\cite{Hasler}.\footnote{This variant representation of $\cN=2$ supersymmetry 
was called the ``three-form multiplet'' in \cite{Hasler}, by analogy with 
its $\cN=1$ counterpart constructed by Gates \cite{Gates}.}

It turns out that the action \eqref{1.2} can be coupled to conformal
supergravity. This requires that the central charge be gauged, and the usual
way this is done is with an off-shell vector multiplet. The locally supersymmetric
version of the action \eqref{1.2}
then corresponds to the bilinear coupling between a linear multiplet
and the vector multiplet that gauges the central charge.\footnote{When
the linear multiplet is independent of the central charge, the
action is just the supersymmetric generalization of the topological $BF$ coupling.}
It is natural to ask how much of the above structure survives in the
presence of supergravity -- and the answer turns out to be all of it!
In the next section, we will demonstrate how to construct a four-form $\Sigma$ and
three-form $H$ in the presence of 4D conformal supergravity with a central
charge and explain how the four-form $\Sigma$ leads to the  linear
multiplet action principle \cite{deWvHVP3}.

\section{The linear multiplet in conformal supergravity}\label{section3}

It is well-known how to couple the linear multiplet (without a central charge)
to 5D $\cN=1$ conformal supergravity
both in superspace \cite{KT-M:5DSugra} and at the component level \cite{KO}.
As our interest is mainly in its 4D manifestation, the most natural line of
attack would be to construct its superform in 5D superspace and then recast
5D superspace as 4D superspace with a central charge. 
However, there is as yet no method to reduce 5D superspace to 4D in the presence of
supergravity; indeed, this has been understood at the component level only
recently \cite{BdeWK},
where it was shown explicitly that off-shell 5D conformal supergravity corresponds to
off-shell 4D conformal supergravity with an additional vector
multiplet. Therefore, instead of performing the reduction of the linear
multiplet directly, we will begin first in four dimensions and consider
the coupling of the linear multiplet to 4D conformal supergravity with a central
charge.

There are several superspace formulations of 4D conformal supergravity, depending
on the choice of the superspace gauge group. The formulation developed in
\cite{KLRT-M1} gauges $\rm SO(3,1) \times SU(2)_R$ and can be derived
\cite{KLRT-M2} from a formulation \cite{Howe} which gauges $\rm SO(3,1) \times U(2)_R$.
Neither of these explicitly gauges dilatations or special superconformal
transformations; rather, both admit a super-Weyl invariance
under which the various connections and torsion superfields transform in a
nonlinear fashion. A more general superspace formulation exists \cite{Butter4D} which gauges
the full superconformal group (the other approaches \cite{Howe} and \cite{KLRT-M1} 
can be obtained from  \cite{Butter4D} by imposing appropriate gauge conditions, see  \cite{Butter4D}  
for more details).\footnote{When enlarging the structure group from
$\rm SU(2)_R$ \cite{KLRT-M1} to $\rm U(2)_R$ \cite{Howe}, the algebra of covariant derivatives becomes
more complicated and practically unsuitable for calculations. One might
think that enlarging the structure group further to the full superconformal
group would make the algebra unmanageable; instead, the algebra
magically simplifies \cite{Butter4D}. This is one of the main advantages
of this formulation.} 
This superspace formulation, which has been called
$\cN=2$ conformal superspace, is convenient to use only when multiplets
and actions transform in a well-defined way under the full superconformal
group. The linear multiplet falls into this class. 

Throughout this paper, we make use of the superspace formulation \cite{Butter4D} for $\cN=2$ 
conformal supergravity. All of our results derived below
can be extended to the other two formulations,
given in  \cite{Howe} and \cite{KLRT-M1}, by performing an appropriate gauge fixing
as described in  \cite{Butter4D}. 

The superspace is described by a supermanifold $\cM^{4|8}$ parametrized by
local bosonic $(x)$ coordinates and local fermionic $(\theta, \bar\theta)$
coordinates $z^M = (x^m, \theta^\mu_\imath, \bar\theta_\dmu^\imath)$. The covariant
derivative $\nabla_A = (\nabla_a, \nabla_\alpha^i, \bar\nabla^\dalpha_i)$ is given by
\begin{align}
\nabla_A = E_A + \hf \Omega_A{}^{ab} M_{ab} + \Phi_A{}^{ij} J_{ij} + \ri \Phi_A Y
	+ B_A \mathbb{D} + \frak{F}_{A}{}^B K_B ~,
\end{align}
with $E_A = E_A{}^M \pa_M$ the vielbein, $\Omega_A$ the spin connection,
$\Phi_A{}^{ij}$ and $\Phi_A$ the $\rm SU(2)_R$ and $\rm U(1)_R$ connections,
$B_A$ the dilatation connection, and $\frak{F}_A{}^B$ the special superconformal
connection.
We may extend the superspace to include a gauged central charge $\Delta$,
$[\Delta, \nabla_A]=0$, which commutes with the superconformal generators,
\begin{align}\label{eq_DeltaGens}
0 = [M_{ab}, \Delta] = [J^{ij}, \Delta] = [Y, \Delta] = [\mathbb D, \Delta] = [K^A, \Delta]~,
\end{align}
by introducing gauge covariant derivatives
\begin{align}
\bnabla_A := \nabla_A + V_A \Delta~,\qquad \D V_A=0 \quad 
\longrightarrow \quad [\Delta, \bnabla_A] = 0~.
\end{align}
The gauge transformation of the connection $V_A$ is 
\bea
\d V_A = - \nabla_A \L \quad \longrightarrow \quad 
\d \bnabla_A = [\L \D, \bnabla_A ]~, \qquad \D \L=0~,
\eea
while the other connections, $E_A{}^M$, $\Omega_A{}^{bc}$, etc.
are inert under the central charge transformation. We require
a tensor superfield $\Psi$ and its central charge descendants, $\D \J, \D^2 \J, \dots$,
to transform covariantly under the central charge
\begin{align}
\delta_\L \Psi = \Lambda \Delta \Psi~, \qquad
\delta_\L \Delta \Psi = \Lambda \Delta^2 \Psi~, \quad \dots ~,
\end{align}
which implies that the central charge gauge parameter should itself
be inert,
\begin{align}
\delta_\Lambda \Delta = [\Lambda \Delta, \Delta] = 0~ \quad\Longleftrightarrow\quad
\Delta \Lambda = 0~.
\end{align}
Provided appropriate constraints are imposed
\cite{GSW}, 
the one-form $V_A$ describes a vector
multiplet whose field strength is the reduced chiral superfield $\cZ$.
For further details and the algebra of covariant derivatives, we
refer the reader to Appendix \ref{app_A} as well as the references
\cite{Butter4D, BN}.

It is possible to interpret the central charge $\Delta$ as a derivative
in a fifth bosonic direction, which can simplify some of the equations
we will encounter. Let $z^\hM$ denote
the coordinates of the superspace $\cM^{4|8} \times \cX$
where $\cM$ is a four-dimensional $\cN=2$ supermanifold parametrized
by coordinates $z^M$ and $\cX$ denotes the central charge space
parametrized by $x^5$. The vielbein on this supermanifold is
given by
\begin{align}\label{eq_ccVielbein}
E_{\hM}{}^\hA =
\begin{pmatrix}
E_M{}^A & -V_M \\
0 & 1
\end{pmatrix}~, \qquad
E_{\hA}{}^\hM =
\begin{pmatrix}
E_A{}^M & V_A \\
0 & 1
\end{pmatrix}~,
\end{align}
and depends only on the coordinates $z^M$ parametrizing $\cM^{4|8}$.
The connections associated with the rest of the superconformal group are completely
localized on $\cM^{4|8}$,
\begin{subequations}\label{eq_ccConnections}
\begin{align}
\Omega_\hA{}^{ab} &= (\Omega_A{}^{ab}, 0)~, \qquad
\Phi_\hA{}^{ij} = (\Phi_A{}^{ij}, 0)~, \qquad \textrm{etc.}~, \\
\pa_5 \Omega_A{}^{ab} &= 0 ~, \qquad \pa_5 \Phi_A{}^{ij} = 0, \qquad \textrm{etc.}
\end{align}
\end{subequations}
This choice for the vielbein and the other connections is preserved
so long as we restrict to $x^5$-independent gauge transformations.
We may then define
\begin{align}\label{eq_ccCovD}
\widehat\bnabla_\hA := E_\hA{}^\hM \pa_{\hM}
	+ \hf \Omega_\hA{}^{ab} M_{ab} + \Phi_\hA{}^{ij} J_{ij} + \ri \Phi_\hA Y
	+ B_\hA \mathbb{D} + \frak{F}_{\hA}{}^B K_B~,
\end{align}
which possesses the algebra
\begin{align} \label{eq_algccCovD}
[\widehat\bnabla_\hA, \widehat\bnabla_\hB\} &= T_{\hA\hB}{}^\hC \widehat\bnabla_\hC
	+ \hf R_{\hA\hB}{}^{cd} M_{cd} + R_{\hA\hB}{}^{kl} J_{kl}
	\eol & \quad
	+ \ri R_{\hA\hB}(Y) Y + R_{\hA\hB} (\mathbb{D}) \mathbb{D} + R_{\hA\hB}{}^C K_C~.
\end{align}
Given the choices we have made for the vielbein and the connections, it is
easy to see that
\begin{align}
\widehat\bnabla_A = \bnabla_A~, \qquad \widehat\bnabla_5 = \pa_5 = \Delta~,
\end{align}
and so the algebra of covariant derivatives \eqref{eq_algccCovD} becomes
\begin{subequations}\label{eq_algCovD}
\begin{align}
[\bm \nabla_A, \bm \nabla_B\} &= T_{AB}{}^C \bm \nabla_C + F_{AB} \Delta
	+ \hf R_{AB}{}^{cd} M_{cd} + R_{AB}{}^{kl} J_{kl}
	\eol & \quad
	+ \ri R_{AB}(Y) Y + R_{AB} (\mathbb{D}) \mathbb{D} + R_{AB}{}^C K_C~, \\
[\Delta, \bm\nabla_A] &= 0~,
\end{align}
\end{subequations}
provided we make the identification $F_{AB} = T_{AB}{}^5$. This leads to
\begin{align}
F = \rd V \quad\Longleftrightarrow\quad
F_{AB} = 2 \bm\nabla_{[A} V_{B\}} - T_{AB}{}^C V_C~,
\end{align}
with the Bianchi identity
\begin{align}
\rd F = 0 \quad\Longleftrightarrow\quad
\bm\nabla_{[A} F_{BC\}} - T_{[AB|}{}^D F_{D|C\}} = 0~.
\end{align}
The algebra \eqref{eq_algCovD} is exactly that described in Appendix \ref{app_A}.
Naturally, the central charge gauge transformation arises from a
diffeomorphism in the $x^5$-direction and must be \emph{independent} of $x^5$
to preserve the form \eqref{eq_ccVielbein} for the vielbein.
It should be kept in
mind that although this superspace is formally five-dimensional, it describes only
4D $\cN=2$ conformal supergravity with a central charge, and \emph{not} 5D conformal
supergravity. We will refer to this superspace as central charge
superspace.\footnote{A related superspace involving a complex
central charge was constructed in \cite{AGHH}.}

We generalize the flat linear multiplet by introducing a closed superspace
four-form $\widehat\Sigma$ in central charge superspace,
\begin{align}\label{eq_SigmaBianchi5}
\hat\rd \widehat\Sigma = 0 \quad \Longleftrightarrow \quad
\widehat\bnabla_{[\hA} \widehat\Sigma_{\hB \hC \hD \hE\}}
	- 2 T_{[\hA \hB|}{}^{\hF} \widehat\Sigma_{\hF| \hC \hD \hE\}} = 0~.
\end{align}
This closed form decomposes into a four-form and a three-form when written
in 4D superspace. Denoting
\begin{align}
\Sigma_{ABCD} = \widehat\Sigma_{ABCD}~, \qquad H_{ABC} = \widehat \Sigma_{5ABC}~,
\end{align}
for a four-form $\Sigma$ and a three-form $H$, the equation
\eqref{eq_SigmaBianchi5} decomposes into two equations,
\begin{subequations}
\begin{align}
\bm\nabla_{[A} \Sigma_{BCDE\}}
	- 2 T_{[A B|}{}^{F} \Sigma_{F| C D E\}}
	&= 2 F_{[A B} H_{C D E\}}~, \\
4 \bm\nabla_{[A} H_{BCD]}
	- 2 T_{[A B|}{}^{E} H_{E| C D\}} &= \Delta \Sigma_{ABCD}~,
\end{align}
\end{subequations}
which may equivalently be written
\begin{align}\label{eq_HSigmaBianchi}
\bm\nabla \Sigma = H \wedge F~, \qquad \bm\nabla H = \Delta \Sigma~,
\end{align}
where $\bnabla := E^A \bnabla_A$ is the covariant exterior derivative
of 4D superspace. The superforms $H$ and $\Sigma$ are required
to transform as scalars under central charge gauge transformations,
\begin{align}
\delta_\Lambda H_{ABC} = \Lambda \Delta H_{ABC}~, \qquad
\delta_\Lambda \Sigma_{ABCD} = \Lambda \Delta \Sigma_{ABCD}~.
\end{align}
Imposing the constraints
\begin{gather}\label{eq_Hconstraints}
H_{\ul{\alpha \beta \gamma}} = H_{\ul{\alpha \beta \dgamma}} = 0~, \qquad
H_a{}_\beta^i{}_\gamma^j = H_a{}_\dbeta^i{}_\dgamma^j = 0~, \qquad
H_a{}_\beta^i{}_\dgamma^j = 2 (\sigma_a)_{\beta \dgamma} \cL^{ij}~, \\
\Sigma_{\ul {\halpha\hbeta\hgamma\hdelta}} = 0~, \qquad
\Sigma_{a \ul{\hbeta \hgamma \hdelta}} = 0~, \qquad
\Sigma_{a b}{}_\alpha^i{}_\dbeta^j = 0~,\label{eq_Sigmaconstraints}
\end{gather}
we find that the superfield $\cL^{ij}$ must be a linear multiplet,
\begin{align}\label{eq_Linear}
\bm \nabla_\alpha^{(i} \cL^{jk)} = \bar{\bm \nabla}_\dalpha^{(i} \cL^{jk)} = 0~.
\end{align}
The remaining components of $H$ are given by
\begin{align}\label{eq_H}
H_{ab}{}_\a^i = - \frac{2 \ri}{3} (\s_{ab})_\a{}^\b \bm \nabla_{\b j} \cL^{ji}~,\qquad
H_{abc} = \frac{\ri}{24} \veps_{abcd} (\ts^d)^{\dalpha\alpha} [\bm \nabla_\a^k, \bar{\bm \nabla}_\ad^l] \cL_{kl}~,
\end{align}
and those of $\Sigma$ are given by
\begin{subequations}\label{eq_Sigma}
\begin{align}
\Sigma_{ab}{}_\a^i{}_\b^j &= 4 \ri (\s_{ab})_{\a\b} \bar{\cZ} \cL^{ij} ~,\\
\Sigma_{abc}{}_\a^i &= - \frac{\ri}{2} \veps_{abcd} (\s^d)_{\a\ad} \bar{\bm \nabla}^\ad_j \bar{\cZ} \cL^{ij} - \frac{\ri}{3} \veps_{abcd} (\s^d)_{\a\ad} \bar{\cZ} \bar{\bm \nabla}^\ad_j \cL^{ij}~, \\
\Sigma_{abcd} &= \frac{1}{24} \veps_{abcd} (\cZ \bm \nabla_{kl} \cL^{kl} + \bar{\cZ} \bar{\bm \nabla}_{kl} \cL^{kl} + 3 \bm \nabla^{kl} \cZ \cL_{kl} \non\\
&\quad \quad + 4 \bm \nabla^{\g k} \cZ \bm \nabla_\g^l \cL_{kl} + 4 \bar{\bm \nabla}_{\gd k} \bar{\cZ} \bar{\bm \nabla}^\gd_l \cL^{kl} )~.
\end{align}
\end{subequations}
These results can be compared with those in the previous section by
setting $\bm\nabla_A \rightarrow D_A$ and $\cZ \rightarrow 1$.

As in the flat case, two special situations are noteworthy.
The first is if $\cL^{ij}$ is
taken to be independent of the central charge, $\Delta \cL^{ij} = 0$,
then $H$ is closed in the usual sense, 
\bea 
\Delta \cL^{ij} = 0 \quad \longrightarrow \quad 
\rd H = 0~,
\label{3.25}
\eea 
and $\cL^{ij}$ becomes a tensor multiplet.
The second situation is if we choose the closed form $\widehat\Sigma$
to be exact,
\begin{align}
\widehat \Sigma = \hat\rd \widehat C \quad\Longleftrightarrow\quad
\widehat\Sigma_{\hA \hB\hC\hD} &= 4 \widehat\bnabla_{[\hA} \widehat C_{\hB\hC\hD\}}
	- 6 T_{[\hA\hB|}{}^\hE \widehat C_{\hE|\hC\hD\}}~.
\label{3.26}
\end{align}
This implies that $H$ and $\Sigma$ are
given in terms of a two-form $B_{AB} = -\widehat C_{5 AB}$ and a three-form $C_{ABC} = \widehat C_{ABC}$,
\begin{subequations}
\begin{align}
H_{ABC} &= 3 \bnabla_{[A} B_{BC\}} - 3 T_{[AB|}{}^D B_{D|C\}} + \Delta C_{ABC}~,\\
\Sigma_{ABCD} &= 4 \bnabla_{[A} C_{BCD\}} - 6 T_{[AB|}{}^E C_{E|CD\}} + 6 F_{[AB}{} B_{CD\}}~,
\end{align}
\end{subequations}
or, equivalently,
\begin{align}
H = \bnabla B + \Delta C~, \qquad \Sigma = \bnabla C + B \wedge F~.
\end{align}
As in the flat case, this leads to a variant representation for the linear
multiplet in 4D where one of its auxiliaries is the divergence of a 
vector. The supergravity generalization of eq. \eqref{eq_LAux}, however, is quite
complicated, so we will not construct it explicitly here.
In the case that the conditions \eqref{3.25} and \eqref{3.26} are imposed simultaneously, 
we obtain a variant realization of the tensor multiplet such that one of its auxiliaries is 
replaced by the field strength of a gauge three-form. We can think of this realization as a 
three-form multiplet in $\cN=2$ conformal supergravity.

Now we would like to interpret $\Sigma_{abcd}$ as (part of)
a supersymmetric Lagrangian. This turns out to be possible using
the so-called ectoplasm formalism \cite{Ectoplasm,GGKS}.
The key element of this approach is a superspace four-form $J$ which is
closed.\footnote{On a usual four-dimensional manifold, any four-form is closed
trivially, but in superspace the condition is nontrivial.} The action constructed
by integrating $J$ over the manifold $\cM$ parametrized by the physical coordinates
$x^m$ turns out to be automaticaly supersymmetric, which we will demonstrate shortly.

In our case, $\Sigma$ is not itself closed, but we may easily construct a related
four-form that is:
\begin{align}
J := \Sigma + V \wedge H~.
\end{align}
It is straightforward to check that $J$ is closed,
\begin{align}
\rd J &= \rd \Sigma + V \wedge \rd H - \rd V \wedge H
	= \bm\nabla \Sigma - V \wedge \Delta\Sigma + V \wedge \bm\nabla H - F \wedge H = 0~,
\end{align}
using eqs. \eqref{eq_HSigmaBianchi}. We can construct a supersymmetric action via the
integration of $J$ over the manifold $\cM$:\footnote{We define the Levi-Civita
tensor with world indices as $\veps^{mnpq} := \veps^{abcd} e_a{}^m e_b{}^n e_c{}^p e_d{}^q$.}
\begin{align}
S = \int_\cM J
	= \int \rd^4x\, e\, (^*J)~, \qquad ^*J = \frac{1}{4!} \veps^{mnpq} J_{mnpq}~.
\end{align}
This action is automatically supersymmetric by virtue of the closure of $J$.
The proof is straightforward. Since supersymmetry is the combination
of a superdiffeomorphism and a gauge transformation, it suffices to show that
the action is invariant separately under superdiffeomorphisms and gauge
transformations. First, we observe that a superdiffeomorphism is a super Lie derivative:
\begin{align}
\delta_\xi J \equiv \cL_\xi J \equiv \imath_\xi \rd J + \rd\imath_\xi J = \rd\imath_\xi J~,
\end{align}
with the last equality following since $J$ is closed. Provided that the manifold
$\cM$ has no boundary, the variation of the action is zero.
Next, we consider gauge transformations. Since $J$ is a scalar under the
superconformal generators (Lorentz, $\rm U(2)_R$, dilatation and special superconformal),
the only nontrivial check involves the central charge gauge transformation.
We note that
\begin{align}
\delta_\Lambda J = \delta_\Lambda \Sigma + \delta_\Lambda V \wedge H + V \wedge \delta_\L H~,
\end{align}
but $\Sigma$ and $H$ both transform covariantly under central charge gauge transformations,
$\delta_\Lambda \Sigma = \Lambda \Delta \Sigma$ and $\delta_\Lambda H = \Lambda \Delta H$, while $V$
transforms as a connection, $\delta_\Lambda V = -\rd \Lambda$. So we find
\begin{align}
\delta_\Lambda J &= \Lambda \Delta \Sigma - \rd\Lambda \wedge H + V \wedge \Lambda \Delta H
	= \rd (\Lambda H)~.
\end{align}
Once again $J$ transforms into an exact form and so the action $S$ is invariant.

We can now give the supersymmetric action explicitly. We identify
\begin{align}
J_{mnpq} = \Sigma_{mnpq} - 4 V_{[m} H_{npq]}
\end{align}
or equivalently,
\begin{align}
^*J = \frac{1}{4!} \veps^{mnpq} \Sigma_{mnpq} - \frac{1}{3!} \veps^{mnpq} V_m H_{npq}~.
\end{align}
The second term is a topological $BF$ coupling; the first term is its supersymmetric
completion and is given by
\begin{align}
\frac{1}{4!} \veps^{mnpq} \Sigma_{mnpq}|
	&= \frac{1}{4!} \veps^{mnpq} E_q{}^D E_p{}^C E_n{}^B E_m{}^A \Sigma_{ABCD}| \non\\
	&= \frac{1}{4!} \veps^{abcd} \Big( \frac{1}{2} \Sigma_{abcd} + 2 \psi_a{}^\a_i \Sigma_\a^i{}_{bcd}
		+ \frac{3}{2} \psi_b{}^\b_j \psi_a{}^\a_i \Sigma_\a^i{}_\b^j{}_{cd} \Big) + \HC \non\\
	&= -\frac{1}{2} F \phi - \frac{1}{2} \chi^\a_i \l_\a^i
	- \frac{1}{16} \ell^{ij} X_{ij}
	+ \frac{\ri}{4} \psi_{\a \ad}{}^\a_i ( 2 \bar{\chi}^{\ad i} \bar{\phi} + \ell^{ij} \bar{\l}^{\ad}_j )
	\eol & \quad
	- \frac{1}{2} (\s^{cd})_{\g \d} \psi_c{}^\g_k \psi_d{}_l^\d \ell^{kl} \bar{\phi} + {\rm c.c.} \ ,
\end{align}
where
\begin{subequations}
\begin{alignat}{2}
\ell^{ij} &:= \cL^{ij}|~, \\
\c_{\a i} &:= \frac{1}{3} \bm \nabla_\a^j \cL_{i j}| ~, &\quad 
	\bar{\c}^{\ad i} &:= \frac{1}{3} \bar{\bm \nabla}^\ad_j \cL^{ij}|~, \\
F &:=  \frac{1}{12} \bm \nabla^{ij} \cL_{ij}|~, &\quad
\bar{F} &:=  \frac{1}{12} \bar{\bm \nabla}^{ij} \cL_{ij}| \ ,
\end{alignat}
\end{subequations}
and
\begin{align}
\phi := \cZ| \ , \quad \l{}_\a^i :=\bm \nabla_\a^i \cZ| \ , \quad X^{ij} := \bm \nabla^{ij} \cZ | \ .
\end{align}
The full action is
\begin{align}\label{eq_LinearAction}
S &= -\frac{1}{2} \int \rd^4x\, e\, \Big(F \phi + \chi^\a_i \l_\a^i
	+ \frac{1}{8} \ell^{ij} X_{ij} + \frac{1}{6} \veps^{mnpq} V_m H_{npq}
	\eol & \quad
	- \frac{\ri}{2} \psi_{\a \ad}{}^\a_i ( 2 \bar{\chi}^{\ad i} \bar{\phi} + \ell^{ij} \bar{\l}^{\ad}_j )
	+(\s^{cd})_{\g \d} \psi_c{}^\g_k \psi_d{}_l^\d \ell^{kl} \bar{\phi} + {\rm c.c.}\Big)~,
\end{align}
which agrees with  \cite{deWvHVP3}. (This action is equivalent to that given in
\cite{BS} up to a gauge-fixing.)
The terms $V_m$ and $H_{npq}$ are understood as the projections of the corresponding
superforms. Up to a normalization factor, this is exactly the supersymmetric action coupling a
linear multiplet to the vector multiplet gauging the central charge.

\section{A deformed linear multiplet}\label{section4}

Now we turn to the main point of our paper: the generalization of the
linear multiplet when the central charge is gauged by a more elaborate
multiplet. We describe first a superspace where the central charge
connection itself transforms under the central charge, reviewing the
construction given recently in \cite{Novak}. Then we reexamine the
structure of the coupled four-form $\Sigma$ and three-form $H$ to
discover a generalized version of the linear multiplet. This naturally
implies a generalized version of the action principle \eqref{eq_LinearAction}.

\subsection{A large vector multiplet}
Until now we have gauged the central charge using a normal $\cN=2$ vector multiplet --
that is, the vector multiplet was inert under the central charge. A generalization
immediately presents itself: we may choose the central charge gauge connection
to no longer be inert under $\Delta$. We identify
\begin{align}\label{eq_defbmnabla}
\bm\nabla_A := \nabla_A + \cV_A \Delta~, \qquad \D \cV_A \neq 0~,
\end{align}
where $\nabla_A$ is the original covariant derivative of conformal supergravity, while
$\cV_A$ is the gauge connection associated with $\Delta$. 
The gauge transformation of $\cV_A$ is 
\bea
\d \cV_A = - \nabla_A \L + \L \D \cV_A \quad 
\longrightarrow \quad 
\d \bnabla_A = [\L \D, \bnabla_A ]~, \qquad \D \L=0~.
\eea
Unlike the gauge one-form $\cV_A$, the gauge parameter is neutral with respect to the central charge. 
As before, the central charge commutes with the other generators, \eqref{eq_DeltaGens},
but because $\Delta \cV_A \neq 0$, we find that $[\Delta, \bnabla_A] \neq 0$.

A five-dimensional interpretation is even more useful now than before.
Again, we take the vielbein of the larger superspace to be
\begin{align}
E_{\hM}{}^\hA =
\begin{pmatrix}
E_M{}^A & -\cV_M \\
0 & 1
\end{pmatrix}~, \qquad
E_{\hA}{}^\hM =
\begin{pmatrix}
E_A{}^M & \cV_A \\
0 & 1
\end{pmatrix}~.
\end{align}
We allow $\cV_A = E_A{}^M \cV_M$ to depend on the fifth bosonic coordinate,
but we take $E_A{}^M$ to be independent of $x^5$ as before.
The connections are given again by \eqref{eq_ccConnections}
and the covariant derivative by \eqref{eq_ccCovD},
leading to
\begin{align}
\widehat\bnabla_\hA = (\bnabla_A, \Delta)~, \qquad \Delta = \pa_5~.
\end{align}
The algebra of covariant derivatives \eqref{eq_algccCovD} now decomposes into
\begin{subequations}
\begin{align}
[\bm \nabla_A, \bm \nabla_B\} &= T_{AB}{}^C \bm \nabla_C + \cF_{AB} \Delta
	+ \hf R_{AB}{}^{cd} M_{cd} + R_{AB}{}^{kl} J_{kl}
	\eol & \quad
	+ \ri R_{AB}(Y) Y + R_{AB} (\mathbb{D}) \mathbb{D} + R_{AB}{}^C K_C~, \\
[\Delta, \bm\nabla_A] &= \cF_{5 A} \Delta~, \label{eq_vCCalg2}
\end{align}
\end{subequations}
provided we make the identifications
\begin{align}
\cF = T^5 \qquad\Longleftrightarrow\qquad \cF_{AB} = T_{AB}{}^5~, \quad \cF_{5A} = T_{5A}{}^5~,
\end{align}
which leads to
\begin{align}
\cF_{AB} = 2 \bm\nabla_{[A} \cV_{B\}} - T_{AB}{}^C \cV_C~, \qquad
\cF_{5A} = \Delta \cV_A = \pa_5 \cV_A~.
\end{align}
The torsion tensor $T^5$ is closed by construction,
\begin{align}
\widehat\bnabla T^5 = 0 \quad\Longleftrightarrow\quad
\widehat\bnabla_{[\hC} T_{\hB \hA\}}{}^5 - T_{[\hC \hB|}{}^\hD T_{\hD |\hA\}}{}^5 = 0~,
\end{align}
where $\widehat\bnabla := E^\hA \widehat\bnabla_\hA$.
This implies similar relations for $\cF$,
\begin{align}\label{eq_dF1}
\bm\nabla_{[A} \cF_{BC\}} - T_{[AB|}{}^D \cF_{D|C\}}
	&= \cF_{[AB|} \cF_{5|C\}}~, \\
2 \bm\nabla_{[A|} \cF_{5|B\}} - T_{AB}{}^D \cF_{5 D} &= \Delta \cF_{AB}~. \label{eq_dF2}
\end{align}
If we introduce the two-form $\cF$ and the one-forms $\cV$ and $\cF_5$, defined by
\begin{align}
\cF = \frac{1}{2} E^B \wedge E^A \cF_{AB}~, \qquad
\cF_5 = E^A \cF_{5A}~, \qquad
\cV = E^A \cV_A~,
\end{align}
then $\cF$ and $\cF_5$ can be written
\begin{align}
\cF = \rd \cV + \Delta \cV \wedge \cV~, \qquad
\cF_5 = \Delta \cV~,
\end{align}
and the Bianchi identities become
\begin{align}
\bnabla \cF \equiv \rd \cF + \Delta \cF \wedge \cV = \cF_5 \wedge \cF~, \qquad
\bnabla \cF_5 \equiv \rd \cF_5 + \Delta \cF_5 \wedge \cV = \Delta \cF~,
\end{align}
with $\bnabla:= E^A \bnabla_A$.

Let us now impose constraints on the field strength $\cF$. In analogy
to the $x^5$-independent case, we take\footnote{Our definition of $M$ differs by a factor 
of $\ri$ from \cite{Novak}.}
\begin{align}
\cF_\alpha^i{}_\beta^j = 2 \veps_{\alpha \beta} \veps^{ij} \bar M~, \qquad
\cF^\dalpha_i{}^\dbeta_j = -2 \veps^{\dalpha \dbeta} \veps_{ij} M~, \qquad
\cF_\alpha^i{}^\dbeta_j = 0~,
\end{align}
where $M$ is a conformally primary superfield of dimension 1 and $U(1)$ charge $-2$.
Analyzing the Bianchi identities, we find that $M$ must obey two constraints,
\begin{gather} \label{eq_MBianchi1}
\bm\nabla_\alpha^{(i} \bar{\bm\nabla}_\dalpha^{j)} \ln \left(\frac{M}{\bar M}\right) = 0~, \\
\bar M \bm\nabla^{ij} \left(\frac{M}{\bar M} \right) = 
M \bar {\bm\nabla}^{ij} \left(\frac{\bar M}{M} \right)~. \label{eq_MBianchi2}
\end{gather}
The remaining components of $\cF$ are then determined to be
\begin{subequations}
\begin{align}
\cF_a{}_\beta^j &= -\frac{\ri}{2} (\sigma_a)_\beta{}^\dalpha \bar M
	\bar{\bm\nabla}_\dalpha^j \ln \left(\frac{\bar M}{M}\right)~, \\
\cF_{ab} &= \frac{1}{8} (\sigma_{ab})^{\alpha \beta} 
	(\bar M \bm\nabla_{\alpha \beta} \Big(\frac{M}{\bar M}\Big) + 4 \bar M W_{\alpha \beta}) + \HC~, \\
\cF_5{}_\alpha^i &= \bm\nabla_\alpha^i \ln \bar M~, \\
\cF_{5a} &= -\frac{\ri}{8} (\sigma_a)_{\alpha \dalpha}
	(\bm\nabla^{\alpha k} \bar{\bm\nabla}_k^\dalpha \ln M
	+ \bar{\bm\nabla}_k^\dalpha \bm\nabla^{\alpha k} \ln \bar M)~.
\end{align}
\end{subequations}
It is straightforward to check that if $\cV_A$ is $x^5$-independent, then
$M$ becomes a reduced chiral superfield $\cZ$.

This large vector multiplet has an interesting feature.
Although $[\Delta, \bm\nabla_A] = \cF_{5A}\Delta$ is nonzero, we can easily see that
\begin{align}
[\bar M \Delta, \bm\nabla_\alpha^i] = 0~, \qquad
[M \Delta, \bar{\bm\nabla}^\dalpha_i] = 0~.
\end{align}

\subsection{Deformed linear multiplet} 

Now let us construct a deformation of the linear multiplet in four
dimensions. The constraints we will impose are quite cumbersome if we
insist on a purely four dimensional superspace interpretation.
In 4D superspace, we take a four-form $\Sigma$ and a three-form $H$ to obey
the constraints
\begin{align}
\bm\nabla \Sigma = H \wedge \cF~, \qquad \bm\nabla H + H \wedge \cF_5 = \Delta \Sigma~,
\end{align}
which can equivalently be written
\begin{subequations}\label{eq_varHSigmaBianchi}
\begin{align}
\bm\nabla_{[A} \Sigma_{BCDE\}}
	- 2 T_{[A B|}{}^{F} \Sigma_{F| C D E\}}
	&=  2 \cF_{[A B} H_{C D E\}}~, \\
4 \bm\nabla_{[A} H_{BCD]}
	- 2 T_{[A B|}{}^{E} H_{E| C D\}}
	+ 4 \cF_{5 [A} H_{BCD\}} &= \Delta \Sigma_{ABCD}~.
\end{align}
\end{subequations}
Central charge superspace offers a more economical way of encoding the
above equations. The superforms $\Sigma$ and $H$ may be placed within a single
superform $\widehat \Sigma$,
\begin{align}
\Sigma_{ABCD} = \widehat \Sigma_{ABCD}~, \qquad H_{ABC} = \widehat\Sigma_{5 A B C}~.
\end{align}
We require $\widehat\Sigma$ to be closed, which amounts to
\begin{align}
0 &= \widehat {\bm\nabla}_{[\hA} \widehat \Sigma_{\hB\hC\hD\hE\}}
	- 2 T_{[\hA \hB|}{}^{\hF} \widehat\Sigma_{\hF| \hC \hD \hE\}}~.
\end{align}
This equation is equivalent to the two equations \eqref{eq_varHSigmaBianchi}.

By fixing some of the lowest components of $\Sigma$ and $H$, one can show that
they are completely specified by a deformed linear multiplet $\cL^{ij}$,
obeying
\begin{align}\label{eq_variantL}
\bnabla_{\alpha}^{(i} (\bar M \cL^{jk)}) = 0~, \qquad
\bar\bnabla_{\dalpha}^{(i} (M \cL^{jk)}) = 0~.
\end{align}
It is useful to introduce tilded derivatives defined as
\begin{align}
\widetilde\bnabla{}_\alpha^i = \bar M^{-1} \bnabla_\alpha^i \bar M~, \qquad
\widetilde{\bar\bnabla}{}^\dalpha_i = M^{-1} \bar\bnabla^\dalpha_i M~, \qquad
\end{align}
so the conditions \eqref{eq_variantL} can be more compactly written
\begin{align}
\widetilde \bnabla{}_{\alpha}^{(i} \cL^{jk)} = 0~, \qquad
\widetilde{\bar\bnabla}{}_{\dalpha}^{(i} \cL^{jk)} = 0~.
\end{align}
The components of the three-form $H_{ABC} = \widehat \Sigma_{5 ABC}$ are
\begin{subequations}\label{eq_varH}
\begin{align}
    H_{\ul{\halpha \hbeta \hgamma}} &= 0~, \qquad
    H_{a}{}_\beta^i{}_\gamma^j = H_{a}{}^\dbeta_i{}^\dgamma_j = 0~, \qquad
    H_{a}{}_\beta^i{}^\dgamma_j = 2 (\sigma_a)_{\beta}{}^\dgamma \cL^i{}_j~, \\
    H_{ab}{}_\alpha^i &= \frac{2\ri}{3} (\sigma_{ab})_\alpha{}^\gamma \widetilde \bnabla{}_\gamma^k \cL_k{}^i~, \quad
    H_{ab}{}^\dalpha_i = \frac{2\ri}{3} (\tilde\sigma_{ab})^\dalpha{}_\dgamma \widetilde{\bar\bnabla}{}^\dgamma_k \cL^k{}_i~,\\
    H_{abc} &= \frac{\ri}{24} \veps_{abcd} (\tilde \sigma^d)^{\dgamma \gamma}
        [\widetilde \bnabla_{\gamma j}, \widetilde{\bar \bnabla}_{\dgamma k}] \cL^{jk}
\end{align}
\end{subequations}
and the components of the four-form $\Sigma_{ABCD} = \widehat\Sigma_{ABCD}$ are
\begin{subequations}\label{eq_varSigma}
\begin{align}
\Sigma_{\ul{\halpha \hbeta \hgamma \hdelta}}
	&= \Sigma_{a \ul{\hbeta \hgamma \hdelta}}
	= \Sigma_{ab}{}_\alpha^i{}^\dbeta_j = 0~, \\
\Sigma_{ab}{}_\alpha^i{}_\beta^j &= 4 \ri (\sigma_{ab})_{\alpha \beta} \bar M \cL^{ij}~, \quad
\Sigma_{ab}{}^\dalpha_i{}^\dbeta_j = - 4 \ri (\tilde\sigma_{ab})^{\dalpha \dbeta} M \cL_{ij}~, \\
\Sigma_{abc}{}_\alpha^i &= - \ri \,\veps_{abcd} (\sigma^d)_{\alpha \dgamma} \left(
	\frac{1}{2} M \bar \bnabla^\dgamma_{k} \Big(\frac{\bar M}{M}\Big) \cL^{ki}
	+ \frac{1}{3} \bar M \widetilde {\bar\bnabla}{}^\dgamma_{k} \cL^{ki} \right)~, \\
\Sigma_{abc}{}^\dalpha_i &= - \ri \,\veps_{abcd} (\tilde\sigma^d)^{\dalpha \gamma} \left(
	\frac{1}{2} \bar M \bnabla_\gamma^{k} \Big(\frac{M}{\bar M}\Big) \cL_{ki}
	+ \frac{1}{3} M \widetilde \bnabla{}_\gamma^{k} \cL_{ki} \right)~, \\
\Sigma_{abcd} &= \frac{1}{24} \veps_{abcd} \left(
	M \widetilde{\bnabla}_{jk} \cL^{jk}
	+ 4 \bar M \bnabla^{\gamma}_j \Big(\frac{M}{\bar M}\Big) \widetilde{\bnabla}_{\gamma k} \cL^{jk}
	+ \frac{3}{2} \bar M \bnabla_{jk} \Big(\frac{M}{\bar M}\Big) \cL^{jk}
	+ \HC
	\right)~.
\end{align}
\end{subequations}

\subsection{Locally supersymmetric action} 
Let us apply the ectoplasm method to the four-form constructed in the previous subsection. 
The superform $\Sigma$ obeys the equation
\begin{align}
\rd \Sigma &= H \wedge \cF - \Delta \Sigma \wedge \cV
	= \rd (H \wedge \cV)~,
\end{align}
so we can introduce the closed four-form
\begin{align}
J:= \Sigma - H \wedge \cV~,
\end{align}
which transforms under a central charge gauge transformation as an exact form,
\begin{align}
\delta J = \Lambda \Delta \Sigma - \Lambda \Delta H \wedge \cV - H \wedge \rd \Lambda
	= \rd (\Lambda H)~.
\end{align}
Then by the argument made in the previous section, we may define an action using
the Lagrangian
\begin{align}
^*J := \frac{1}{4!} \veps^{mnpq} J_{mnpq} =
	\frac{1}{4!} \veps^{mnpq} \Sigma_{mnpq} - \frac{1}{3!} \veps^{mnpq} \cV_m H_{npq}~.
\end{align}
This is naturally supersymmetric and gauge-invariant. The explicit action is
easy to construct once we note the similarities between eqs.
\eqref{eq_Hconstraints},  \eqref{eq_H} and eqs. \eqref{eq_varH}
and also between eqs. \eqref{eq_Sigmaconstraints}, \eqref{eq_Sigma} and 
eqs. \eqref{eq_varSigma}. We need only make the identifications\begin{subequations}
\begin{alignat}{2}
\ell^{ij} &:= \cL^{ij}|~, \\
\c_{\a i} &:= \frac{1}{3} \widetilde \bnabla{}_\a^j \cL_{i j}| ~, &\quad 
	\bar{\c}^{\ad i} &:= \frac{1}{3} \widetilde {\bar\bnabla}{}^\ad_j \cL^{ij}|~, \\
F &:=  \frac{1}{12} \widetilde \bnabla{}^{ij} \cL_{ij}|~, &\quad
\bar{F} &:=  \frac{1}{12} \widetilde{\bar\bnabla}{}^{ij} \cL_{ij}| \ ,
\end{alignat}
\end{subequations}
and
\begin{align}
\phi := M| \ , \quad \l{}_\a^i := \bar M \bnabla_\a^i \Big(\frac{M}{\bar M}\Big)| \ , \quad
X^{ij} := \bar M \bnabla^{ij} \Big(\frac{M}{\bar M}\Big)| \ .
\end{align}
The full action is then formally identical to the action \eqref{eq_LinearAction}
\begin{align}\label{eq_varLinearAction}
S &= -\hf \int \rd^4x\, e\, \Big(
F \phi 
+\chi^\a_i \l_\a^i
	+ \frac{1}{8} \ell^{ij} X_{ij} + \frac{1}{6} \veps^{mnpq} \cV_m H_{npq}
	\eol & \quad
	- \frac{\ri}{2} \psi_{\a \ad}{}^\a_i ( 2 \bar{\chi}^{\ad i} \bar{\phi} + \ell^{ij} \bar{\l}^{\ad}_j )
	+(\s^{cd})_{\g \d} \psi_c{}^\g_k \psi_d{}_l^\d \ell^{kl} \bar{\phi} + {\rm c.c.}\Big)~.
\end{align}
The difference at the component level is that the supersymmetry transformation
rules of the large vector and deformed linear multiplets have been altered.

Of course, it is easily seen that if the large vector multiplet $M$ is restricted to
be chiral, $\bar\bnabla^\dalpha_i M = 0$, then it reduces to the usual vector
multiplet.\footnote{The chirality condition on $M$ actually implies that $M$ is
independent of the central charge from consistency of the anticommutator
$\{\bar\bnabla^\dalpha_i, \bar \bnabla^\dbeta_j\} M = 0$.} Similarly,
the conditions on the deformed linear multiplet \eqref{eq_variantL}
reduce in this case to the usual constraints \eqref{eq_Linear}
for a linear multiplet.

\section{Applications and discussion}

Until now we have made use of the superfield $M$ with little comment as to its
physical content. It should be apparent that relative to the usual vector multiplet $\cZ$,
the multiplet $M$ is quite enormous; and because it possesses nontrivial
dilatation and $\rm U(1)_R$ weights, we cannot consistently eliminate either
its modulus or its phase. Nevertheless, there are several ways we might attempt
to reduce it.

The simplest choice is (of course) to take $M$ to be independent
of the central charge, which amounts to choosing $M = \cZ$ for some vector
multiplet and reducing all of the structure in section \ref{section4} to
that of section \ref{section3}.
A less trivial alternative is merely to isolate
the $x^5$-dependence into either the modulus or the phase of $M$.
The first choice is to take
\begin{align}
\pa_5 (M \bar M) \neq 0~, \qquad 
\pa_5 (M / \bar M) = 0~.
\end{align}
Examining the Bianchi identity \eqref{eq_MBianchi1}, we see that it is solved by
\begin{align}
M / \bar M = \Phi / \bar \F~,
\end{align}
for some chiral superfield $\Phi$. But \eqref{eq_MBianchi2} then gives
$\nabla^{ij} \Phi = \bar\nabla^{ij} \bar \Phi$ and so $\Phi$ is
a reduced chiral multiplet, $\Phi = \cZ$. The Bianchi identities
then tell us \emph{absolutely nothing} about the modulus of $M$.

Now consider the second choice,
\begin{align}
\pa_5 (M \bar M) = 0~, \qquad 
\pa_5 (M / \bar M) \neq 0~.
\end{align}
The Bianchi identities tell us little about
$M \bar M$, which is some real superfield with dilatation weight two.
But because $M \bar M$ is $x^5$-independent,
we may treat it as a conformal supergravity compensator.
In this light, the most natural choice would seem to be
$M \bar M = \cZ \bar \cZ$ for a vector multiplet $\cZ$.
Making this choice for the modulus of $M$,
we identify the phase by setting
\begin{align}
M = -\ri \cZ \re^{-\ri L}
\end{align}
for some real superfield $L$ that depends on the central charge.\footnote{The overall
choice of phase of $M$ can be changed by redefining $L$. The choice made here
matches that used in \cite{Novak}.} The Bianchi identities \eqref{eq_MBianchi1}
and \eqref{eq_MBianchi2} then become
\begin{align}\label{eq_varVT1}
\bnabla_\alpha^{(i} \bar\bnabla_\dbeta^{j)} L = 0~, \qquad
\re^{\ri L} \bnabla^{ij} (\cZ \re^{-2 \ri L}) = - \re^{-\ri L} \bar\bnabla^{ij} (\bar\cZ \re^{2 \ri L})~.
\end{align}
These equations are (some of) the constraints that define the variant
vector-tensor multiplet \cite{Novak}.

The variant vector-tensor multiplet has been defined recently in supergravity \cite{Novak}.
It is a generalization (both to supergravity and with more general couplings
to vector multiplets) of a multiplet introduced first by Theis \cite{Theis1, Theis2}.
The simplest version involves introducing the additional constraint \cite{Novak} 
\begin{align}\label{eq_varVT2}
\re^{-\ri L} \bnabla^{ij} (\cZ \re^{2 \ri L}) = - \re^{\ri L} \bar\bnabla^{ij} (\bar\cZ \re^{-2 \ri L})~.
\end{align}
The superfield $L$ obeying \eqref{eq_varVT1} and \eqref{eq_varVT2}
describes the variant vector-tensor multiplet. Its Lagrangian
is constructed from a generalized linear multiplet $\cL^{ij}$ given by \cite{Novak}
\begin{align}
\cL^{ij} = \frac{\ri}{2} \re^{-\ri L} \bnabla^{ij} (\cZ L \re^{2 \ri L})
	- \re^{\ri L} \cZ \bnabla^{\alpha i} L \bnabla_\alpha^j L
	- \frac{1}{4} \re^{\ri L} \bnabla^{ij} \cZ + \HC
\end{align}
which can be shown to obey the constraints
\begin{align}
0 = \re^{-\ri L} \bnabla_\alpha^{(i} (\re^{\ri L} \cL^{jk)}) = \widetilde\bnabla{}_\alpha^{(i} \cL^{jk)}~, \qquad
0 = \re^{\ri L} \bar\bnabla_\dalpha^{(i} (\re^{-\ri L} \cL^{jk)}) = \widetilde{\bar\bnabla}{}_\dalpha^{(i} \cL^{jk)}~, \qquad
\end{align}
We refer the reader to \cite{Novak} for a full discussion.

An alternative possibility is to use the variant vector-tensor multiplet
(or some other central charge multiplet $M$) to
construct a new action involving a {\it massless} Fayet-Sohnius hypermultiplet.
Let us suppose $q_i$ is a superfield obeying the constraints
\begin{align}
\bnabla_\alpha^{(i} q^{j)} = \bar \bnabla{}_\dbeta^{(i} q^{j)} = 0
\end{align}
and similarly for its conjugate $\bar q^i:=(q_i)^*$.
We can introduce a composite variant linear multiplet
\begin{align}
\cL^{ij} = \frac{1}{2} \bar q^{(i} \stackrel{\longleftrightarrow}{\Delta} q^{j)}~.
\end{align}
By construction, $\cL^{ij}$ obeys
\begin{align}
\widetilde\bnabla{}_\alpha^{(i} \cL^{jk)} = \widetilde\bnabla{}_\dbeta^{(i} \cL^{jk)} = 0
\end{align}
and its action may be constructed directly using \eqref{eq_varLinearAction}.
As with the usual Fayet-Sohnius hypermultiplet, this multiplet
has the equation of motion $\Delta q_i = 0$ and so 
the on-shell hypermultiplet decouples from the large vector multiplet.\footnote{In the case that the central charge is gauged using a standard vector multiplet,  the hypermultiplet Lagrangian 
can include a mass term,  $\cL^{ij} = \frac{1}{2} \bar q^{(i} \stackrel{\leftrightarrow}{\Delta} q^{j)}
+\ri \, m \, \bar q^{(i}  q^{j)}$ , with $m$ a real mass parameter. 
No mass term is allowed if a large vector multiplet is used.  }

It has recently been shown at the component level \cite{BdeWK} that
5D $\cN=1$ conformal supergravity can be dimensionally reduced off-shell
to 4D $\cN=2$ conformal supergravity coupled to a vector multiplet.
One expects that this component construction can be repeated at the
superfield level and thereby connect 5D $\cN=1$ superspace directly to
the
central charge superspace we considered in section \ref{section3}.
A natural question to ask is whether the more general central charge
structure
described in section \ref{section4}, involving a large vector multiplet,
has any significance from a 5D point of view. In particular, can one construct
actions in 5D which preferentially reduce in 4D so that the central charge
multiplet retains $x^5$-dependence?

We are aware of no examples, but there is one interesting possibility.
It was pointed out in \cite{KL} that the nonlinear vector-tensor multiplet
has a simple 5D origin, at least in flat superspace. It was noted recently by
two of us (DB and JN) \cite{BN} that the generalization of the
vector-tensor
multiplet to conformal supergravity \cite{Claus1, Claus2} can also be
interpreted as arising from a certain 5D action. In both of these
situations,
the central charge is gauged by the usual vector multiplet.
However, the variant vector-tensor multiplet \cite{Theis1, Theis2,
Novak}
\emph{itself} gauges the central charge, and so the central charge
multiplet must retain $x^5$-dependence. Should this variant VT multiplet
possess a 5D origin, it would provide just such an example.

We conclude this paper with a final comment. 
Within the superconformal tensor calculus, the two main types of locally supersymmetric
actions are: (i) the chiral action; and (ii) the linear multiplet action. 
The ectoplasm construction for the chiral action was given in  \cite{Gates:2009xt}. 
The case of the linear multiplet action has been worked out in the present paper. 
\\

\noindent
{\bf Acknowledgements:}\\
The work  of DB and SMK  is supported in part by the Australian Research Council.
The work of JN is supported by an Australian Postgraduate Award.

\appendix

\section{Conformal supergravity in 4D $\cN=2$ superspace}\label{app_A}
In this appendix, we briefly summarize the algebra of $\cN=2$ conformal
superspace \cite{Butter4D} as reformulated in \cite{BN}.

\subsection{$\cN=2$ conformal superspace}\label{app_CS}
The covariant derivative $\nabla_A = (\nabla_a, \nabla_\alpha^i, \bar\nabla^\dalpha_i)$
is given by
\begin{align}
\nabla_A &= E_A + \hf \Omega_A{}^{ab} M_{ab} + \Phi_A{}^{ij} J_{ij} + \ri \Phi_A Y
	+ B_A \mathbb{D} + \frak{F}_{A}{}^B K_B \eol
	&= E_A + \Omega_A{}^{\b\g} M_{\b\g} + \bar{\Omega}_A{}^{\bd\gd} \bar{M}_{\bd\gd}
	+ \Phi_A{}^{ij} J_{ij} + \ri \Phi_A Y + B_A \mathbb{D} + \frak{F}_{A}{}^B K_B ~.
\end{align}
Here $E_A = E_A{}^M \pa_M$ is the supervielbein, $\Omega_A{}^{ab}$ is the spin connection,
and $\Phi_A{}^{ij}$ and $\Phi_A$ are the $\rm SU(2)_R$ and $\rm U(1)_R$ connections,
respectively. In addition, we have a dilatation connection $B_A$ and a special
superconformal connection $\frak F_A{}^B$.

The Lorentz generators $M_{ab}$ obey
\begin{align}
[M_{ab}, \nabla_c ] &= 2 \eta_{c [a} \nabla_{b]}~, \quad
[M_{ab}, \nabla_\a^i] = (\s_{ab})_\a{}^\b \nabla_\b^i ~,\quad 
[M_{ab}, \bar\nabla^\ad_i] = (\tilde{\s}_{ab})^\ad{}_\bd \bar\nabla^\bd_i~.
\end{align}
As usual, they may be decomposed into left-handed and right-handed generators
\begin{subequations}
\bea
M_{\alpha \beta} &=& \frac{1}{2} (\sigma^{ab})_{\alpha \beta} M_{ab}~, \qquad
\bar M_{\dalpha \dbeta} = -\frac{1}{2} (\tilde\sigma^{ab})_{\dalpha \dbeta} M_{ab}~, \quad \\
M_{ab} &= &(\sigma^{ab})_{\alpha \beta} M_{\alpha \beta} - (\tilde\sigma^{ab})_{\dalpha \dbeta} \bar M_{\dalpha \dbeta}~,
\eea
\end{subequations}
which act only on undotted and dotted indices, respectively
\begin{align}
[M_{\alpha \beta}, \nabla_\gamma^i] = \ve_{\gamma (\alpha} \nabla_{\beta)}^i~, \qquad
[\bar M_{\dalpha \dbeta}, \bar\nabla_{\dgamma i}] = \ve_{\dgamma (\dalpha} \bar\nabla_{\dbeta) i}~.
\end{align}
The $\rm SU(2)_R$, $\rm U(1)_R$ and dilatation generators obey
\begin{align}
[J_{ij}, \nabla_\a^k] &= - \d^k_{(i} \nabla_{\a j)} ~,\quad
[J_{ij}, \bar\nabla^\ad_k] = - \ve_{k (i} \bar\nabla^{\ad}_{j)}~, \non \\
[Y, \nabla_\a^i] &= \nabla_\a^i ~,\quad [Y, \bar\nabla^\ad_i] = - \bar\nabla^\ad_i~,  \non \\
[\mathbb{D}, \nabla_a] &= \nabla_a ~, \quad
[\mathbb{D}, \nabla_\a^i] = \hf \nabla_\a^i ~, \quad
[\mathbb{D}, \bar\nabla^\ad_i] = \hf \bar\nabla^\ad_i ~.
\end{align}
The special superconformal generators $K^A = (K^a, S^\alpha_i, \bar S_\dalpha^i)$
transform in the obvious way under Lorentz and $\rm SU(2)_R$ generators,
\begin{align}
[M_{ab}, K_c] &= 2 \eta_{c [a} K_{b]} ~, \quad
[M_{ab} , S^\g_i] = - (\s_{ab})_\b{}^\g S^\b_i ~, \quad
[M_{ab} , \bar S_\gd^i] = - (\s_{ab})^\bd{}_\gd \bar S_\bd^i~, \non \\
[J_{ij}, S^\g_k] &= - \ve_{k (i} S^\g_{j)} ~, \quad
[J_{ij}, \bar{S}^k_\gd] = - \d^k_{(i} \bar{S}_{\gd j)}~,
\end{align}
and carry opposite $\rm U(1)_R$ and dilatation weight to $\nabla_A$:
\begin{align}
[Y, S^\a_i] &= - S^\a_i ~, \quad
[Y, \bar{S}^i_\ad] = \bar{S}^i_\ad~, \non \\
[\mathbb{D}, K_a] &= - K_a ~, \quad
[\mathbb{D}, S^\a_i] = - \hf S^\a_i ~, \quad
[\mathbb{D}, \bar{S}_\ad^i] = - \hf \bar{S}_\ad^i ~.
\end{align}
Among themselves, the generators $K^A$ obey the algebra
\begin{align}
\{ S^\a_i , \bar{S}^j_\ad \} &= 2 \ri \d^j_i (\s^a)^\a{}_\ad K_a~.
\end{align}
with all the other (anti-)commutators vanishing.

Finally, the algebra of $K^A$ with $\nabla_B$ is given by
\begin{align}
[K^a, \nabla_b] &= 2 \delta^a_b \mathbb{D} + 2 M^{a}{}_b ~,\non \\
\{ S^\a_i , \nabla_\b^j \} &= 2 \d^j_i \d^\a_\b \mathbb{D} - 4 \d^j_i M^\a{}_\b 
- \d^j_i \d^\a_\b Y + 4 \d^\a_\b J_i{}^j ~,\non \\
\{ \bar{S}^i_\ad , \bar{\nabla}^\bd_j \} &= 2 \d^i_j \d^\bd_\ad \mathbb{D} 
+ 4 \d^i_j \bar{M}_\ad{}^\bd + \d^i_j \d_\ad^\bd Y - 4 \d_\ad^\bd J^i{}_j ~,\non \\
[K^a, \nabla_\b^j] &= -\ri (\s^a)_\b{}^\bd \bar{S}_\bd^j \ , \quad [K^a, \bar{\nabla}^\bd_j] = 
-\ri ({\s}^a)^\bd{}_\b S^\b_j ~, \non \\
[S^\a_i , \nabla_b] &= \ri (\s_b)^\a{}_\bd \bar{\nabla}^\bd_i \ , \quad [\bar{S}^i_\ad , \nabla_b] = 
\ri ({\s}_b)_\ad{}^\b \nabla_\b^i \ ,
\end{align}
where all other (anti-)commutations vanish.

The algebra of covariant derivatives has the form
\begin{align}
[\nabla_A, \nabla_B\} 
         &= T_{AB}{}^C \nabla_C + \hf R_{AB}{}^{cd} M_{cd} + R_{AB}{}^{kl} J_{kl}
	\eol & \quad
	+ \ri R_{AB}(Y) Y + R_{AB} (\mathbb{D}) \mathbb{D} + R_{AB}{}^C K_C ~.
\end{align}
We impose constraints on the curvatures appearing on the right-hand side to
reproduce the component structure of conformal supergravity \cite{Butter4D}.
The resulting algebra is given by
\begin{subequations}\label{CSGAlgebra}
\begin{align}
\{ \nabla_\a^i , \nabla_\b^j \} &= 2 \ve^{ij} \ve_{\a\b} \bar{W}_{\gd\dd} \bar{M}^{\gd\dd} + \hf \ve^{ij} \ve_{\a\b} \bar{\nabla}_{\gd k} \bar{W}^{\gd\dd} \bar{S}^k_\dd - \hf \ve^{ij} \ve_{\a\b} \nabla_{\g\dd} \bar{W}^\dd{}_\gd K^{\g \gd}~, \\
\{ \bar{\nabla}^\ad_i , \bar{\nabla}^\bd_j \} &= - 2 \ve_{ij} \ve^{\ad\bd} W^{\g\d} M_{\g\d} + \frac{1}{2} \ve_{ij} \ve^{\ad\bd} \nabla^{\g k} W_{\g\d} S^\d_k - \frac{1}{2} \ve_{ij} \ve^{\ad\bd} \nabla^{\g\gd} W_{\g}{}^\d K_{\d \gd}~, \\
\{ \nabla_\a^i , \bar{\nabla}^\bd_j \} &= - 2 \ri \d_j^i \nabla_\a{}^\bd~, \\
[\nabla_{\a\ad} , \nabla_\b^i ] &= - \ri \ve_{\a\b} \bar{W}_{\ad\bd} \bar{\nabla}^{\bd i} - \frac{\ri}{2} \ve_{\a\b} \bar{\nabla}^{\bd i} \bar{W}_{\ad\bd} \mathbb{D} - \frac{\ri}{4} \ve_{\a\b} \bar{\nabla}^{\bd i} \bar{W}_{\ad\bd} Y + \ri \ve_{\a\b} \bar{\nabla}^\bd_j \bar{W}_{\ad\bd} J^{ij}
	\eol & \quad
	- \ri \ve_{\a\b} \bar{\nabla}_\bd^i \bar{W}_{\gd\ad} \bar{M}^{\bd \gd} - \frac{\ri}{4} \ve_{\a\b} \bar{\nabla}_\ad^i \bar{\nabla}^\bd_k \bar{W}_{\bd\gd} \bar{S}^{\gd k} + \frac{1}{2} \ve_{\a\b} \nabla^{\g \bd} \bar{W}_{\ad\bd} S^i_\g
	\eol & \quad
	+ \frac{\ri}{4} \ve_{\a\b} \bar{\nabla}_\ad^i \nabla^\g{}_\gd \bar{W}^{\gd \bd} K_{\g \bd}~, \\
[ \nabla_{\a\ad} , \bar{\nabla}^\bd_i ] &=  \ri \d^\bd_\ad W_{\a\b} \nabla^{\b}_i + \frac{\ri}{2} \d^\bd_\ad \nabla^{\b}_i W_{\a\b} \mathbb{D} - \frac{\ri}{4} \d^\bd_\ad \nabla^{\b}_i W_{\a\b} Y + \ri \d^\bd_\ad \nabla^{\b j} W_{\a\b} J_{ij}
	\eol & \quad
	+ \ri \d^\bd_\ad \nabla^{\b}_i W^\g{}_\a M_{\b\g} + \frac{\ri}{4} \d^\bd_\ad \nabla_{\a i} \nabla^{\b j} W_\b{}^\g S_{\g j} - \hf \d^\bd_\ad \nabla^\b{}_\gd W_{\a\b} \bar{S}^{\gd}_i
	\eol & \quad
	+ \frac{\ri}{4} \d^\bd_\ad \nabla_{\a i} \nabla^\g{}_\gd W_{\b\g} K^{\b\gd} ~.
\end{align}
\end{subequations}
We have not given the algebra of two vector covariant derivatives since it
may straightforwardly be derived as a consequence of the above (anti)-commutators.
The result is given in \cite{Butter4D}.

The curvatures are characterized by a single complex superfield $W_{\alpha \beta}$,
which is the superconformal Weyl tensor. It is symmetric ($W_{\a\b} = W_{\b\a}$),
superconformally primary ($K_A W_{\a\b} = 0$), chiral ($\bar{\nabla}^\ad_i W_{\b\g} = 0$),
and obeys the Bianchi identity
\begin{align}
\nabla_{\a\b} W^{\a\b} &= \bar{\nabla}^{\ad\bd} \bar{W}_{\ad\bd} ~,
\end{align}
where we introduce the notation
\begin{align}
\nabla_{\a\b} := \nabla_{(\a}^k \nabla_{\b) k} \ , \quad \bar{\nabla}^{\ad\bd} := \bar\nabla^{(\ad}_k \bar\nabla^{\bd) k} \ .
\end{align}

\subsection{Conformal supergravity and central charge} \label{CSCC}
As in, {\it e.g.} \cite{BN} we can introduce a central charge gauged by
an off-shell vector multiplet.  First, we introduce a modified covariant derivative:
\be \bm \nabla_A := \nabla_A + V_A \D \ ,
\ee
where $V_A(z)$ is the gauge connection and $\D$ is a real central charge. We assume
that the central charge commutes with the modified covariant derivative,
$[\Delta, \bm \nabla_A] = 0$, which allows us to treat it completely analogously
to a $\rm U(1)$ generator as far as the algebra is concerned.

The curvature tensors are given by
\begin{align}
[\bm \nabla_A, \bm \nabla_B\} &=
	T_{AB}{}^C \bm \nabla_C
	+ F_{AB} \Delta
	+ \hf R_{AB}{}^{cd} M_{cd} + R_{AB}{}^{kl} J_{kl}
	\eol & \quad
	+ \ri R_{AB}(Y) Y + R_{AB} (\mathbb{D}) \mathbb{D} + R_{AB}{}^C K_C~.
\end{align}
We then impose the constraints for the vector multiplet, using
$\cZ$ to denote the corresponding abelian field strength,
\begin{subequations}
\begin{align}
F_\a^i{}_\b^j &= 2 \ve_{\a\b} \ve^{ij} \bar{\cZ} \ , \quad
F^\ad_i{}^\bd_j = -2 \ve^{\ad\bd} \ve_{ij} \cZ \ , \qquad F_\a^i{}^\bd_j = 0 ~,\\
F_a{}_\b^j &= -\frac{\ri}{2} (\s_a)_\b{}^\gd \bar{\bm \nabla}_\gd^j \bar{\cZ} \ , \qquad 
F_a{}^\bd_j = \frac{\ri}{2} (\s_a)_\g{}^\bd \bm \nabla^\g_j \cZ \ , \\
F_{ab} &= \frac{1}{8} (\s_{ab})_{\a\b} ( \bm\nabla^{\a \b} \cZ + 4 W^{\a\b} \bar{\cZ})
	- \frac{1}{8} (\tilde{\s}_{ab})_{\ad\bd}  (\bar{\bm \nabla}^{\ad \bd} \bar{\cZ} + 4 \bar{W}^{\ad\bd} \cZ) \ ,
\end{align}
\end{subequations}
where $\cZ$ is a reduced chiral primary superfield with dimension 1 and $\rm U(1)$ weight $-2$
\begin{align}
K_A \cZ &= 0 \ , \quad \mathbb{D} \cZ = \cZ \ , \quad Y \cZ = -2 \cZ \ , \non\\
\bar{\nabla}_{\ad}^i \cZ &= 0~,\qquad
\nabla^{ij} \cZ = \bar{\nabla}^{ij} \bar{\cZ} ~.
\end{align}
Our normalization for $\cZ$ has been chosen so that in the flat limit taking
$\cZ\rightarrow 1$ allows the identification of the central charge curvature $F$
with the five-dimensional torsion tensor $T^5$.

\section{$\cN=1$ $BF$ coupling via ectoplasm}
In this appendix, we briefly discuss how to use the ectoplasm method to construct
the $BF$ action in $\cN=1$ conformal supergravity corresponding to
\begin{align}\label{eq_BFN1}
S_{\rm BF} = \int \rd^4x \, \rd^4\q\, E \, L V
\end{align}
where $V$ is a vector multiplet prepotential and $L$ is a real linear multiplet
obeying
\begin{align}
\nabla^2 L = \bar\nabla^2 L = 0~.
\end{align}
We work in $\cN=1$ conformal superspace \cite{Butter:2009cp}
but use the Lorentz conventions consistent with the rest of the
paper.\footnote{The standard formulation of $\cN=1$ conformal
supergravity \cite{Howe}
can be obtained from this formulation by an appropriate
gauge-fixing \cite{Butter:2009cp, Butter:2011vg}.}

Let $\Sigma$ be a superspace four-form obeying the equation
\begin{align}\label{eq_dSigma1}
\rd \Sigma + F \wedge H = 0
\end{align}
for closed two-form $F$ and closed three-form $H$. The field strength $F$ is given by
\begin{subequations}
\begin{align}
F_{\hat\beta \hat\alpha} &= 0~, \qquad
F_{\beta a} = (\sigma^a)_{\beta \dbeta} \bar W^\dbeta~, \qquad
F_{\dbeta a} = (\sigma^a)_{\beta \dbeta} W^\beta~, \\
F_{ba} &= \frac{\ri}{2} (\sigma_{ba})^{\alpha \beta} \nabla_\beta W_\alpha
	- \frac{\ri}{2} (\tilde\sigma_{ba})_{\dalpha \dbeta} \bar\nabla^\dbeta \bar W^\dalpha~,
\end{align}
\end{subequations}
where $W_\alpha$ is a reduced chiral spinor obeying
$\nabla^\alpha W_\alpha = \bar\nabla_\dalpha \bar W^\dalpha$.
The three-form $H$ is given by
\begin{subequations}
\begin{align}
H_{\hat \gamma \hat\beta \hat\alpha} &= 0~, \qquad
H_{\gamma \beta a} = H_{\dgamma \dbeta a} = 0~, \qquad
H_{\gamma \dbeta a}= 2\ri (\sigma_a)_{\gamma \dbeta} L~, \\
H_{\gamma ba} &= -2 (\sigma_{ba})_\gamma{}^\delta \nabla_\delta L~, \qquad
H_{\dgamma ba} = -2 (\tilde\sigma_{ba})_{\dgamma \ddelta} \bar\nabla^\ddelta L~, \\
H_{cba} &= -\frac{1}{4} \veps_{dcba} (\tilde\sigma^d)^{\dalpha \alpha} [\nabla_\alpha, \bar\nabla_\dalpha] L~.
\end{align}
\end{subequations}
If we constrain certain components of the four-form $\Sigma$ to be zero, we find
\begin{subequations}
\begin{align}
\Sigma_{\hat\delta \hat\gamma \hat\beta A} &= 0~, \\
\Sigma_{\delta \gamma b a} &= \frac{1}{16} (\sigma_{ba})_{\delta \gamma} \bar Y~,\qquad
\Sigma^{\ddelta \dgamma}{}_{b a} = \frac{1}{16} (\tilde\sigma_{ba})^{\ddelta \dgamma} Y~,\qquad
\Sigma_{\delta \dgamma b a} = 0~, \\
\Sigma_{\delta c b a} &= -\frac{1}{16} \veps_{dcba} (\sigma^d)_{\delta \ddelta}
	(\bar \nabla^\ddelta \bar Y - 16 \ri \,\bar W^\ddelta L)~, \\
\Sigma^{\ddelta}{}_{c b a} &= +\frac{1}{16} \veps_{dcba} (\tilde\sigma^d)^{\ddelta \delta}
	(\nabla_\delta Y + 16 \ri \,W_\delta L)~, \\
\Sigma_{dcba} &= \veps_{dcba} \Big(
	\frac{\ri}{64} (\nabla^2 Y - \bar \nabla^2 \bar Y)
	- W^\alpha \nabla_\alpha L - \bar W_\dalpha \bar\nabla^\dalpha L
	- \frac{1}{2} L \nabla^\alpha W_\alpha \Big)~,
\end{align}
\end{subequations}
where $Y$ is a chiral superfield. The $Y$-dependent pieces of $\Sigma$
correspond to a closed four-form $\Sigma_c$, which was constructed
in \cite{GGKS} (see also \cite{Gates:2009xt}) and generates a chiral
action in $\cN=1$ supergravity.
This closed four-form $\Sigma_c$ coincides with the field strength of a gauge
three-form in the case that $Y = \bar\nabla^2 X$ with $X = \bar X$
\cite{Gates, BGG}.

Now we introduce a closed four-form $J$
\begin{align}
J := \Sigma + A \wedge H~, \qquad
\rd J = \rd \Sigma - \rd A \wedge H = 0~.
\end{align}
The four-form $J_{mnpq}$,
\begin{align}
J_{mnpq} = \Sigma_{mnpq} - 4 A_{[m} H_{npq]}~,
\end{align}
then gives a supersymmetric four-form action. Neglecting gravitinos, we find
\begin{align}
\frac{1}{4!} \veps^{mnpq} J_{mnpq} &= 
	\frac{\ri}{64} (\nabla^2 Y - \bar \nabla^2 \bar Y)
	- W^\alpha \nabla_\alpha L - \bar W_\dalpha \bar\nabla^\dalpha L
	- \frac{1}{2} L \nabla^\alpha W_\alpha
	\eol & \quad
	- \frac{1}{4} A^{\dalpha \alpha} [\nabla_\alpha, \bar \nabla_\dalpha] L
	+ \mathcal O(\psi)~.
\end{align}
If we further constrain $Y=0$, this is exactly the component Lagrangian
for the action \eqref{eq_BFN1} under the identifications
$W_\alpha = -\dfrac{1}{8} \bar\nabla^2 \nabla_\alpha V$ and
$A_{\alpha \dalpha} = -\dfrac{1}{4} [\nabla_\alpha, \bar\nabla_\dalpha] V$.

\begin{footnotesize}

\end{footnotesize}

\end{document}